\documentclass[a4,10pt]{article}
\usepackage{multirow}
\usepackage{graphics}
\usepackage{array}
\usepackage{enumerate}
\usepackage{graphics}
\usepackage{subfigure}

\newcommand{\ket}[1]{\ensuremath{\left| #1 \right\rangle}}

\newcommand{\br}[1]{\ensuremath{\left\langle #1 \right.}}
\newcommand{\bra}[1]{\ensuremath{\left. \br{#1} \right|}}
\newcommand{\bk}[2]{\br{{#1}}\ket{{#2}}}
\newcommand{\kb}[2]{\ket{{#1}}\bra{{#2}}}

\newcommand{\proj}[1]{\kb{{#1}}{{#1}}}

\newcommand{\mean}[1]{\ensuremath{\left\langle {#1} \right\rangle}}

\newcommand{\mod}[1]{\ensuremath{\left| {#1} \right|}}
\newcommand{\magn}[1]{\ensuremath{\mod{#1}^2}}
\newcommand{\dirdel}[2]{\delta\left({#1}-{#2}\right)}

\newcommand{\eqnwrap}{\nonumber \\ &&}

\begin{document}
\title{Detectability, Invasiveness and the Quantum Three Box Paradox}
\author{O.~J.~E.~Maroney \\
Faculty of Philosophy, University of Oxford\footnote{Mailing address:  Wolfson College, Linton Road, Oxford, OX2 6UD, United Kingdom}\\
owen.maroney@philosophy.ox.ac.uk}
\date{\today}


\maketitle

\begin{abstract}
Quantum pre- and post-selection (PPS) paradoxes occur when counterfactual inferences are made about different measurements that might have been performed, between two measurements that are actually performed.  The 3 box paradox is the paradigm example of such a paradox, where a ball is placed in one of three boxes and it is inferred that it would have been found, with certainty, both in box 1 and in box 2 had either box been opened on their own.  Precisely what is at stake in PPS paradoxes has been unclear, and classical models have been suggested which are supposed to mimic the essential features of the problem.  We show that the essential difference between the classical and quantum pre- and post-selection effects lies in the fact that for a quantum PPS paradox to occur the intervening measurement, had it been performed, would need to be invasive but non-detectable.  This invasiveness is required even for null result measurements.  While some quasi-classical features (such as non-contextuality and macrorealism) are compatible with PPS paradoxes, it seems no fully classical model of the 3 box paradox is possible.
\end{abstract}

\section{Introduction}
The quantum 3 Box Paradox\cite{AV1991} effect has now been experimentally confirmed in a number of different contexts\cite{RLS2004,KSY+2011} with the most up-to-date tests\cite{GMR+2012} violating a measure of classicality by over seven standard deviations.  As the paradigm example of a quantum pre- and post-selection (PPS) paradox there has been much discussion of its meaning.  Questions raised include the validity of the counterfactual use of the ABL rule\cite{SS1993,Cohen1995,Kastner2003}, connections to quantum contextuality proofs and measurement disturbance\cite{LS2005a,LS2005b} and whether classical systems can simulate the essential properties of a PPS paradox\cite{Kirkpatrick2003,LS2005b}.

Ravon and Vaidman\cite{RV2007} have rejected these classical comparisons, on the basis that they introduce a measurement disturbance.  They argue that the essence of the 3 Box Paradox is that in the classical case there is no reason to suppose the measurement could introduce such a disturbance.

This paper analyses the role played by measurement disturbances in PPS paradoxes and shows precisely what is non-classical about the 3 Box Paradox.  The ontic model framework (introduced in \cite{Spekkens2005a}) will be used to show how measurement disturbance plays a novel role in PPS paradoxes: the existence of a true PPS paradox requires a measurement disturbance which is necessarily \textit{invasive}, but \textit{non-detectable}.

We will show:
\begin{itemize}
\item The Kirkpatrick\cite{Kirkpatrick2003} and Leifer and Spekkens\cite{LS2005b} examples are of a different kind to the 3 Box Paradox.  They involve detectable measurement disturbances that are not present in the 3 Box Paradox, so cannot reproduce the same statistics.  The absence of such detectable disturbances are essential for the 3 Box Paradox, and to any true PPS paradox.
\item However, the key feature of a PPS paradox, which is not present in any simpler system than the 3 Box Paradox, is that although the measurement has no detectable effect, it must in fact disturb the actual state of the system.  Although this is similar to a result obtained by Leifer and Spekkens, they proved only that a non-contextual model must introduce measurement disturbances.  Here we will show any model of a PPS paradox must involve non-detectable disturbances.
\item We will then show how this kind of disturbance can be related to violations of the Leggett Garg Inequality\cite{LG1985}, which may also be argued to test for a form of measurement disturbance.  We show that a PPS paradox holds if, and only if, a Leggett Garg Inequality is violated.  This is surprising, as in 2-dimensional Hilbert spaces the Leggett Garg Inequality can only be violated for detectable measurement disturbances.  This analysis has already been applied in experimental tests of the 3 Box Paradox\cite{GMR+2012}, showing large violations of this classicality measure.
\item Finally explicit examples will be constructed of models which can reproduce the 3 Box Paradox results exactly, while still satisfying certain classical conditions such as macrorealism or non-contextual value definiteness.
\end{itemize}

\section{The 3 Box Paradox}
\subsection{The adversarial game}\label{ss:adversarial}
Alice proposes a game to Bob.  She has three indistinguishable boxes, which she puts in three positions. She puts the a ball in the box in position 3, throws a blanket over the boxes and under cover of the blanket shuffles the boxes around. The blanket is removed and Bob is allowed to look in either the box in position 1 or position 2.  Alice is blindfolded for this, so she is unaware which box Bob looks in (an umpire is present to ensure Bob does not cheat).  Alice then shuffles the boxes around again under cover of her blanket, and looks in the box in position 3.  She wins if, whenever she sees a ball in box 3, Bob also saw the ball in the box he opened.  She offers Bob better than fifty-fifty odds that she will get it right. If she does not see a ball, no bets are placed.

Bob is understandably suspicious.  Perhaps there is some mark being left by the act of opening the box?  To convince Bob this is not the case, Alice makes four suggestions:
\begin{enumerate}
\item Alice gets the umpire to verify that no matter which box Bob checks, the relative frequency with which the ball ends up in position 3 is the same as when no box is checked.
\item Alice promises this: if Bob chooses, he can instead look twice in succession, using the same measurement or the two different measurements.  If Bob finds the ball changes boxes or finds two balls, then he wins immediately.  Otherwise he loses immediately.
\item Alice allows Bob to check the ball location by other means, such as by connecting a spring to the box and seeing if the box is lifted by a force above the weight of the box but below that of the box and ball together.  An empty box is lifted but a full box should be undisturbed.
\item Finally Bob gets to test that he can put the ball in any one of the boxes, perform his different measurements, and then perform any test he likes to see if he can detect any effect of the measurement (he can't).
\end{enumerate}
Bob thinks about this and decides to play the game.  He can't see any way Alice can know which box Bob checked (he's right - Alice doesn't know which box Bob checked).  He reasons that, provided he chooses the boxes to check randomly, Alice can't do better than a fifty-fifty guess whether Bob saw the ball, however she places it.  Still, Alice proceeds to beat him on every round that bets are placed and cleans up.

\subsection{The quantum game}
The state $\ket{i}$ represents the ball being in box $i$.  Initially prepare the system to be in state $\ket{3}$.  Alice's initial shuffle is any unitary which satisfies
\[
U_I \ket{3}=\frac{1}{\sqrt{3}}\left(\ket{1}+\ket{2}+\ket{3}\right)
\].

Three possibilities are allowed for the intervening measurement
 \begin{itemize}
 \item $M_1$: A projective measurement onto $\proj{1}, \frac{1}{2}\left(\proj{2}+\proj{3}\right)$, with outcomes $1$ and $\neg 1$;
 \item $M_2$: A projection measurement onto $\proj{2}, \frac{1}{2}\left(\proj{1}+\proj{3}\right)$, with outcomes $2$ and $\neg 2$.
 \item $N$: Do nothing;
\end{itemize}

After the intervening measurement, Alice's final shuffle is any unitary which satisfies
\[
U_F\frac{1}{\sqrt{3}}\left(\ket{1}+\ket{2}-\ket{3}\right)=\ket{3}
\]
Then for the final measurement perform $M_A$, Alice performs a projection measurement onto $\proj{3}, \frac{1}{2}\left(\proj{1}+\proj{2}\right)$, with outcomes $A$ and $\neg A$.  (We will simplify the later analysis by merging the effects of $U_F$ into $M_A$.)

The three sequences give the statistics:
 \[
 \begin{array}{|c||c|}\hline
    M_1 &
     \begin{array}{cc}
        P_{M_1}(1,A)=1/9 & P_{M_1}(\neg 1,A)=0 \\ P_{M_1}(1, \neg A)=2/9 & P_{M_1}(\neg 1, \neg A)=2/3
     \end{array} \\ \hline
    M_2 &
     \begin{array}{cc}
        P_{M_2}(2,A)=1/9 & P_{M_2}(\neg 2,A)=0 \\ P_{M_2}(2, \neg A)=2/9 & P_{M_2}(\neg 2, \neg A)=2/3
     \end{array} \\ \hline
    N & P_N(A)=1/9 \\ \hline
 \end{array}
\]
from which it can be seen $P_{M_1}(1|A)=P_{M_2}(2|A)=1$

Using these results Alice is certain to win.  So why was Bob tempted to play this game?  Ravon and Vaidman argue that it is something to do with Bob believing that his measurement does not disturb the system.  They argue that in attempts to classically model pre- and post-selection paradoxes, the intervening measurement leaves a mark that makes the post selection impossible ie. when Box 1 is opened, if the ball is \textit{not} observed, then there is some record left of this which prevents the ball ending up in Box 3 for Alice's measurement.  In the case of the 3 Box Paradox, they argue there is no reason to suppose that a classical measurement can disturb the system in this way.
\section{Measurement disturbance in PPS Paradoxes}
\subsection{Non-detectable measurements}
We now highlight an important property of a PPS paradox, absent from the discussion of\cite{Kirkpatrick2003,LS2005b,RV2007}: that Bob's measurements should be non-detectable by Alice.

Bob's measurement $M_i$ is a non-detectable measurement (NDM) for Alice, if and only if
\begin{equation}
P_N(A)=\sum_j P_{M_i}(A,Q_{i,j})
\end{equation}
where $Q_{i,j}$ are the possible outcomes of the $M_i$ measurement.  Alice can gain no information about what measurement Bob performed (or if Bob even performed a measurement) from the statistics of her measurement outcome.

This might seem a reasonable requirement, in itself.  After all, in some of the adversarial games considered (including the 3 Box Paradox) it is not hard to see that if Alice were to have information about what measurement Bob performed, then she could win at the adversarial game without needing to resort to quantum theory.  Although such situations allow Alice to win, there is no special mystery to how.

Perhaps more importantly, Sharp and Shanks\cite{SS1993}, and Cohen\cite{Cohen1995} demonstrated that attempting to combine post-selective inferences when Bob's measurements disturb Alice's measurement results, will in general lead to inconsistent probabilistic predictions.

A simple example of the problem can be given.  Suppose Bob performs a measurement $M$, with a particular outcome $Q$, and then Alice performs measurement $M_A$, with outcomes $A$ and $\neg A$. We can calculate the post-selective inferences $P_M(Q|A)=P_M(Q,A)/P_M(A)$ and $P_M(Q|\neg A)=P_M(Q,\neg A)/P_M(\neg A)$.

Let us assume that these post-selective inferences are valid even if Bob did not make an intervening measurement.  When Alice observes outcome $A$, she infers Bob would have observed $Q$ with probability $P_M(Q|A)$, had he actually made the measurement.  Similarly if she observes outcome $\neg A$, she infers Bob would have observed $Q$ with probability $P_M(Q|\neg A)$, had he actually made the measurement.

Now consider if these inferences are indeed valid, when Bob does not make the measurement, and Alice then observes the probabilities $P_N(A)$ and $P_N(\neg A)$.  Alice is led to calculate that, had Bob actually made the measurement, he would have observed $Q$ with probability $P(Q)=P_M(Q|A)P_N(A)+P_M(Q|\neg A)P_N(\neg A)$.  This is plainly inconsistent with Alice's knowledge that, had Bob actually made the measurement, he would have observed $Q$ with probability $P(Q)=P_M(Q|A)P_M(A)+P_M(Q|\neg A)P_M(\neg A)$.  The only way the combination of inferences could be consistent with Alice's knowledge is if $P_N(A)=P_M(A)$ and $P_N(\neg A)=P_M(\neg A)$: in other words, if Bob's measurement is non-detectable by Alice.


As we show in the appendix, none of the classical models of pre- and post-selection paradoxes presented in the literature\cite{Kirkpatrick2003,LS2005b,RV2007} are non-detectable.  All involve intervening measurements which change the statistics of the final post-selection measurement.  This is not the case in the 3 Box Paradox.

This leaves open the question as to whether a classical model can, in fact, reproduce the statistics of the 3 Box Paradox, in full, and whether the intervening measurements must disturb the system\footnote{Leifer and Spekkens have argued that a non-contextual model must involve measurement disturbance.  However, they do not demonstrate a contextual model must involve disturbance, nor do they give an example of a non-contextual model that reproduces the 3 Box Paradox statistics.}, even though this disturbance cannot be detected.

\subsection{Invasiveness without detectability}\label{s:disturb}
Any given experimental arrangement is characterised by a \textit{preparation} process, $E$, and a \textit{measurement} process, $M$, with distinct outcomes $Q$.  Operationally, this is characterised by a probability $P_{(E,M)}(Q)$.  In the case of quantum theory $P_{(E,M)}(Q)=\magn{\bk{Q}{E}}$.

The ontic state of the system represents the actual state of the world.  This might simply be taken to be the quantum wavefunction for the system, or might represent additional `hidden' variables or elements of reality.  A non-invasive measurement (NIM) is a measurement which does not disturb the ontic state of the system.

we apply the ontic model framework\cite{Spekkens2005a,Rudolph2006,HS2007,HR2007} to prove that, even though the intervening measurement in the 3 Box Paradox is NDM, it still disturbs the ontic state of the system and cannot be NIM.
\begin{enumerate}
\item A preparation process $E$ produces a probabilistic distribution $\mu_E(\lambda)$ over the ontic states $\lambda$. Any convex sum $\mu(\lambda)=\sum w_{E} \mu_{E}(\lambda)$ ($\sum w_{E}=1$, $ w_{E} \geq 0$), is also a valid preparation.

    NB. $\int_\lambda \mu_E(\lambda)=1$.
\item A measurement $M$ is represented by an outcome function, giving the probability of outcome $Q$ occurring, conditional upon the actual ontic state of the system: $\xi_M(Q|\lambda)$.

    NB. $\sum_Q \xi_M(Q| \lambda) =1$
\item The operational probabilities must be recovered through the formula:
 \begin{equation}
 P_{(E,M)}(Q)= \int d\lambda_0  \mu_E(\lambda_0)\xi_M(Q|\lambda_0)
 \end{equation} \label{c:ontic}
\item The probability of a disturbance of the ontic state by the measurement $M$, with outcome $Q$ is given by: $\gamma_M(\lambda_1|Q,\lambda_0)$.

    NB. $\int_{\lambda_1} \gamma_M(\lambda_1|Q,\lambda_0)=1$\label{c:disturb}

\item After the measurement, with outcome $Q$, the new preparation state will be:
    \begin{equation}
    \mu_Q(\lambda)= \int d\lambda_0\mu_E(\lambda_0)\xi_M(Q|\lambda_0)\gamma_M(\lambda|Q,\lambda_0)
    \end{equation}
\end{enumerate}

\textbf{Non-invasive measurability:}  A measurement which has no effect on the ontic state is called a Non-Invasive Measurement (NIM).  It is represented by
\begin{equation}
\gamma_M(\lambda_1|Q,\lambda_0)=\delta(\lambda_1-\lambda_0)
\end{equation}

We now demonstrate the following: the 3 Box Paradox necessarily involves invasive measurements of the ontic state.  The application of the ontic model formalism quickly gives:
\begin{eqnarray}
P_{M_1}(1,A) &=&\int d\lambda_0 d\lambda_1 \mu(\lambda_0)\xi_{M_1}(A|\lambda_0)\gamma_{M_1}(\lambda_1|1,\lambda_0)\xi_{M_A}(A|\lambda_1) \\
P_{M_2}(2,A) &=&\int d\lambda_0 d\lambda_1 \mu(\lambda_0)\xi_{M_2}(2|\lambda_0)\gamma_{M_2}(\lambda_1|2,\lambda_0)\xi_{M_A}(A|\lambda_1) \\
P_{N}(A) &=&\int d\lambda_0  \mu(\lambda_0)\xi_{M_A}(A|\lambda_0) \\
P_{M_1}(A,1)+P_{M_2}(A,2) &=&
 \int d\lambda_0 d\lambda_1 \mu(\lambda_0)\xi_{M_A}(A|\lambda_1)
  \left[ \xi_{M_2}(2|\lambda_0)\gamma_{M_2}(\lambda_1|2,\lambda_0) \right. \eqnwrap
  \left. +\xi_{M_1}(1|\lambda_0)\gamma_{M_1}(\lambda_1|1,\lambda_0) \right]
\end{eqnarray}
We will now assume NIM and show that no PPS paradox can arise.

Consider the overlap between the functions $\xi_{M_2}(2|\lambda)$ and $\xi_{M_1}(1|\lambda)$.  Are there available ontic states $\mu(\lambda)>0$ such that
$\xi_{M_2}(2|\lambda)\xi_{M_1}(1|\lambda)\neq 0$?

If $M_1$ and $M_2$ are NIM, then a measurement of $M_1$ followed by $M_2$ would give the result:
\begin{eqnarray}
P_{M_2,M_1}(2,1) &=&\int d\lambda_0 d\lambda_1 \mu(\lambda_0)\xi_{M_1}(1|\lambda_0)\gamma_{M_1}(\lambda_1|1,\lambda_0)\xi_{M_2}(2|\lambda_1) \nonumber \\
&=&\int d\lambda_0 \mu(\lambda_0)\xi_{M_1}(1|\lambda_0)\xi_{M_2}(2|\lambda_0)
\end{eqnarray}
If there is a non-zero overlap, then $P_{M_2,M_1}(2,1)>0$.  But this means Bob could open Box 1, see a ball inside, then open Box 2 and see a second ball!  Bob would clearly cry ``foul'' at this point!  After all, Alice simply putting a ball in both boxes is a very easy way for her to win, and involves no paradox at all.  We take it that no parties to the discussion would regard this as an acceptable classical explanation of the 3 Box Paradox.  We require $P_{M_2,M_1}(2,1)=0$.   Hence the 3 Box Paradox requires that, if $M_1$ and $M_2$ are NIM, then $\xi_{M_2}(2|\lambda)\xi_{M_1}(1|\lambda)=0$.

Now $\xi_{M_1}(1|\lambda)\leq 1$ and $\xi_{M_2}(2|\lambda)\leq 1$ which together with $\xi_{M_2}(2|\lambda)\xi_{M_1}(1|\lambda)=0$ gives
\begin{equation}
\xi_{M_2}(2|\lambda)+\xi_{M_1}(1|\lambda) \leq 1
\end{equation}
It follows that for NIM measurements, where $\gamma_{M_1}(\lambda_1|1,\lambda_0) =\gamma_{M_2}(\lambda_1|2,\lambda_0) =\delta\left(\lambda_1-\lambda_0\right)$:
\begin{eqnarray}
P_{M_1}(A,1)+P_{M_2}(A,2) &=&
    \int d\lambda_0 d\lambda_1 \mu(\lambda_0)\xi_{M_A}(A|\lambda_1)
        \left[ \xi_{M_2}(2|\lambda_0)\gamma_{M_2}(\lambda_1|2,\lambda_0) \right. \eqnwrap
        \left. +\xi_{M_1}(1|\lambda_0)\gamma_{M_1}(\lambda_1|1,\lambda_0) \right] \nonumber \\
  &=& \int d\lambda_0 \mu(\lambda_0)\xi_{M_A}(A|\lambda_0)
        \left[ \xi_{M_2}(2|\lambda_0)+\xi_{M_1}(1|\lambda_0) \right] \nonumber \\
  &\leq& \int d\lambda_0 \mu(\lambda_0)\xi_{M_A}(A|\lambda_0)=P_{N}(A)
\end{eqnarray}
So NIM implies
\begin{equation}
P_{M_1}(A,1)+P_{M_2}(A,2) \leq P_{N}(A) \label{eq:nopps}
\end{equation}

But the 3 Box Paradox occurs precisely because $P_{M_1}(A,1)+P_{M_2}(A,2) > P_{N}(A)$. As $P_{N}(A)=P_{M_1}(A)=P_{M_2}(A)$, it is simple to rewrite this as $P_{M_1}(1|A)+P_{M_2}(2|A) > 1$.  This is precisely the condition that allows Alice to offer fifty-fifty odds to Bob, yet still expect to win the adversarial game on average.

Hence we conclude there are no possible ontic models for the 3 Box Paradox for which $M_1$ and $M_2$ are both NIM, despite the fact that the $M_1$ and $M_2$ measurements are both NDM.
\section{The Leggett-Garg Inequality}\label{s:lgi}
We have seen that there are two ideas that are necessary for Bob to believe that he has a fair chance at the adversarial game: that the ball is always in one and only one box; that measurement does not disturb the system.  These assumptions are essentially the same assumptions that have been discussed extensively in the context of the Leggett Garg Inequality, under the names macrorealism and non-invasive measurability:
\begin{quote}
\begin{enumerate}
\item Macrorealism per se. A macroscopic system which has available to it two or more
macroscopically distinct states is at any given time in a definite one of those states.
\item Non-invasive measurability. It is possible in principle to determine which of these
states the system is in without any effect on the state itself or on the subsequent
system dynamics.
\end{enumerate}\cite{LG1985}
\end{quote}

The Leggett Garg Inequality has so far been studied exclusively in the context of 2-dimensional Hilbert spaces.  In \cite{MT2012} it is argued that in these situations a Leggett Garg Inequality can only be broken if NDM is violated.  Here we have argued that a true PPS paradox requires NDM is not violated.

However, given the clear similarity between the conditions for a PPS paradox and the conditions for the violation of a Leggett Garg Inequality, it is interesting to ask if there is a connection.  We will now show that, in the case of 3-dimensional Hilbert spaces, it is possible to violate a Leggett Garg Inequality without violating NDM.  A PPS paradox is possible, and Alice can expect to win at the adversarial game, if and only if this Leggett Garg Inequality is violated.

\subsection{Macrorealism}
Macrorealism: The ball is always in one box, and only in one box.  There may be a probability distribution over which box the ball is in, but this must be understood strictly as some form of epistemic uncertainty. In the language of ontic models, we can express this as saying that any preparation $\mu(\lambda)$ is of the form
\begin{equation}
\mu(\lambda)=p_1 \nu_1(\lambda)+p_2 \nu_2(\lambda)+p_3 \nu_3(\lambda) \label{eq:macrorealism}
\end{equation}
where $\nu_i(\lambda)>0$ only if $\xi_M(i|\lambda)=1$ for all measurements $M$ which include the outcome $i$.  In other words, $\nu_1(\lambda)$ is a distribution over ontic states which are certain to be found in Box 1 whenever they are looked for, and similarly for $\nu_2(\lambda)$ and Box 2 etc. Such a model is non-contextually outcome definite for the macrorealist basis\footnote{It does not immediately run into problems with the Kochen Specker theorem, however, as it only requires this to hold for a single basis.}.

\subsection{Non-invasiveness}
For macrorealist theories, there are two ways non-invasiveness can enter into the analysis, that are weaker than NIM, but still able to produce the Leggett Garg Inequality.
\begin{itemize}
\item NIM1. It is possible to determine if the ball is in Box 1 without disturbing the ball when it is actually found in Box 1 (carefully test if the weight of the box is greater than an empty box, for example).

According to NIM1, after we have opened Box 1, and observed that the ball is in the box, the post-measurement preparation state is $\nu_1(\lambda)$.

This means $P_{M_1}(A|1)=\int d\lambda_0 \nu_1(\lambda_0)\xi_{M_A}(A|\lambda_0)$.  Similar statements hold for Box 2 and $M_2$.

\item NIM2. It is possible to determine if the ball is in Box 1 without disturbing the ball when it is \textit{not} actually found in Box 1 (just open Box 1 and don't touch the other boxes, for example).

According to NIM2, after we have opened Box 1, and observed that the ball is not in the box, the post-measurement preparation state is $\frac{p_2\nu_2(\lambda)+p_3\nu_3(\lambda)}{p_2+p_3}$.

This means $P_{M_1}(A|\neg 1)=\int d\lambda_0
\frac{p_2\nu_2(\lambda)+p_3\nu_3(\lambda)}{p_2+p_3}
\xi_{M_A}(A|\lambda_0)$.  Similar statements hold for Box 2 and $M_2$.
\end{itemize}
NIM2 is a statement about \textit{null result} measurements and intuitively might seem more forceful than NIM1.  Our previous result shows at least one of them is false.  In fact, either of these lead to contradictions with the 3 Box Paradox, so both must be false.

First we note that
\begin{equation}
P_{N}(A)=\int d\lambda_0 (p_1\nu_1(\lambda_0)+p_2\nu_2(\lambda_0)+p_3\nu_3(\lambda_0))\xi_{M_A}(A|\lambda_0)
\end{equation}

NIM1 implies straightforwardly that
\begin{eqnarray}
P_{M_1}(A|1)&=&\int d\lambda_0 \nu_1(\lambda_0)\xi_{M_A}(A|\lambda_0) \nonumber \\
P_{M_2}(A|2)&=&\int d\lambda_0 \nu_2(\lambda_0)\xi_{M_A}(A|\lambda_0)
\end{eqnarray}
As $\int d\lambda_0 p_3 \nu_3(\lambda_0)\xi_{M_A}(A|\lambda_0) \geq 0$ it follows
\[
P_{N}(A) \geq p_1 P_{M_1}(A|1)+p_2 P_{M_2}(A|2)
\]

NIM2, on the other hand, implies:
\begin{eqnarray}
P_{M_1}(A|\neg 1)&=&\int d\lambda_0 \frac{p_2\nu_2(\lambda)+p_3\nu_3(\lambda)}{p_2+p_3} \xi_{M_A}(A|\lambda_0)  \nonumber \\
P_{M_2}(A|\neg 2)&=&\int d\lambda_0 \frac{p_1\nu_1(\lambda)+p_3\nu_3(\lambda)}{p_1+p_3} \xi_{M_A}(A|\lambda_0)
\end{eqnarray}

Now we have
\begin{equation}
(p_2+p_3) P_{M_1}(A|\neg 1)+(p_1+p_3) P_{M_2}(A|\neg 2) \geq P_{N}(A) \geq p_1 P_{M_1}(A|1)+p_2 P_{M_2}(A|2)
\end{equation}

As $P_{M_1}(A|\neg 1)=P_{M_2}(A|\neg 2)=0$, $p_1 P_{M_1}(A|1)=p_2 P_{M_2}(A|2)=1/9$ and $P_{N}(A)=1/9$, this would give $0 \geq 1/9 \geq 2/9$ and we have our contradictions (see also\cite{SPS2008} for a related analysis using path integrals).

\subsection{Deriving the Inequality}

This can now be cast into the terms of a Leggett Garg Inequality violation.  If we assign the value $Q=+1$ when the ball is in Box 1 or Box 2, and $Q=-1$ when the ball is in Box 3, the possible sequences of values are:

{\center

\begin{tabular}{|c|c|c||c|}
\hline  $Q_1$ & $Q_2$ & $Q_3$ & $Q$ \\
\hline \hline
\multirow{6}{*}{+1}
  & Box 1: -1 & \multirow{3}{*}{A: +1} & -1 \\
  & Box 2: -1 & & -1 \\
  & Box 3: +1 & & +3 \\ 
  & Box 1: -1 & \multirow{3}{*}{$\neg A$: -1} & -1 \\
  & Box 2: -1 & & -1 \\
  & Box 3: +1 & & -1 \\
 \hline
\end{tabular}

}
\vspace{.1in}

with $Q=Q_1 Q_2 + Q_2 Q_3 + Q_1 Q_3$.  Any probability distribution over these possible outcomes will give $-1 \leq \mean{Q} \leq 3 $.  This is the Leggett Garg Inequality.

Clearly it is not possible to determine $Q_2$ directly with the measurements available to Alice and Bob. However, with the additional assumption of either NIM1 or NIM2, we can calculate a value for $\mean{Q}$ from the observed data to which a macrorealist would be committed, and obtain a contradiction.

We use macrorealism to assume that, even when Bob performs no measurement, there was a matter of fact as to which Box was occupied:
\begin{equation}
P_{N}(A,1)=\int d\lambda_0 p_1 \nu_1(\lambda_0)\xi_{M_A}(A|\lambda_0)
\end{equation}and similarly for $P_{N}(A,2)$ $P_{N}(\neg A,1)$, etc. and substitute these into $\mean{Q}$.

\begin{equation}
\mean{Q}=3P_{N}(A,3)-\left(
        P_{N}(A,1)+P_{N}(\neg A,1)+P_{N}(A,2)+P_{N}(\neg A,2)+P_{N}(\neg A,3)
            \right)
\end{equation}

We now need to use NIM1 or NIM2 to be able to recast this in terms of the probabilities actually observed in Alice and Bob's measurements:
 \[
 \begin{array}{|c||c|}\hline
    M_1 &
     \begin{array}{cc}
        P_{M_1}(1,A) & P_{M_1}(\neg 1,A) \\ P_{M_1}(1, \neg A) & P_{M_1}(\neg 1, \neg A)
     \end{array} \\ \hline
    M_2 &
     \begin{array}{cc}
        P_{M_2}(2,A) & P_{M_2}(\neg 2,A) \\ P_{M_2}(2, \neg A) & P_{M_2}(\neg 2, \neg A)
     \end{array} \\ \hline
    N & P_N(A)\\ \hline
 \end{array}
\]
\subsubsection{NIM1}
If we assume NIM1, we have
\begin{equation}
P_{N}(A,1)=\int d\lambda_0 p_1 \nu_1(\lambda_0)\xi_{M_A}(A|\lambda_0)=P_{M_1}(A,1)
\end{equation}
and similarly $P_{N}(A,2)=P_{M_2}(A,2)$.  From this $P_{N}(A,3)=P_{N}(A)-P_{M_1}(A,1)-P_{M_2}(A,2)$. For $\neg A$, the results $P_{N}(\neg A,1)=P_{M_1}(\neg A,1)$, $P_{N}(\neg A,2)=P_{M_2}(\neg A,2)$ and $P_{N}(\neg A,3)=P_{N}(\neg A)-P_{M_1}(\neg A,1)-P_{M_2}(\neg A,2)$ hold.

\subsubsection{NIM2}
If we assume NIM2, we must use
\begin{equation}
P_{N}(A,2)+P_{N}(A,3)=
\int d\lambda_0 \left(p_2 \nu_2(\lambda_0)+p_3 \nu_3(\lambda_0)\right)\xi_{M_A}(A|\lambda_0)
=P_{M_1}(A,\neg 1)
 \end{equation}
and similarly $P_{M_2}(A,\neg 2)=P_{N}(A,1)+P_{N}(A,3)$, to get
\begin{eqnarray}
P_{N}(A)    &=& P_{N}(A,1)    +P_{N}(A,2)    +P_{N}(A,3) \nonumber \\
P_{M_1}(A) &=& P_{M1}(A,1) +P_{N}(A,2)    +P_{N}(A,3)  \nonumber \\
P_{M_2}(A) &=& P_{N}(A,1)    +P_{M_2}(A,2) +P_{N}(A,3)
\end{eqnarray}
As $P_{N}(A)=P_{M_1}(A)=P_{M_2}(A)$ it follows $P_{N}(A,1)=P_{M_1}(A,1)$ etc. and so we have the same results as for NIM1.

\subsubsection{Leggett Garg Inequality}
Substituting out all terms not directly observed for those we can measure, we obtain an expression that can be tested against experimental data:
\begin{eqnarray}
\mean{Q_{OBS}}
    &=&-\left(
        P_{M_1}(A,1)+P_{M_1}(\neg A,1)+P_{M_2}(A,2)+P_{M_2}(\neg A,2)
            +P_{N}(\neg A) \right. \eqnwrap
            \left. -P_{M_1}(\neg A,1)-P_{M_2}(\neg A,2) \right)
        +3\left(P_{N}(A)-P_{M_1}(A,1)-P_{M_2}(A,2)\right) \nonumber \\
    &=& 4(P_{N}(A)-P_{M_1}(A,1)-P_{M_2}(A,2))-1 \label{eq:lgicond}
\end{eqnarray}
The Leggett Garg Inequality can be violated if, and only if $P_{N}(A)-P_{M_1}(A,1)-P_{M_2}(A,2)<0$.  This can be rewritten:
\begin{equation}
P_{M_1}(1|A)+P_{M_2}(2|A)>1
\end{equation}
Once again this is the condition for which Alice might offer reasonable seeming odds to Bob, and yet still be sure of winning on average.  This completes the proof that a PPS paradox occurs, and Alice can expect to win her adversarial game, if and only if a related Leggett Garg inequality is violated.

The 3 Box Paradox gives $P_{M_1}(1|A)+P_{M_2}(2|A)=2$ and $P_{N}(A)=1/9$.  This gives
\[
\mean{Q_{OBS}}=-\frac{13}{9}<-1
\]
Experimental realisations of the 3 Box Paradox\cite{GMR+2012} have demonstrated this violation of the Leggett Garg Inequality by over 7 standard deviations.

\subsection{Macrorealism and Non-detectable measurements}

In discussion of the Leggett Garg Inequality (in particular \cite{Clifton1990,BGG1994,MT2012}) it has been argued NIM is too strong a condition to rule out macrorealism, as it is sufficient to derive the Leggett Garg Inequality by itself without assuming macrorealism.   The analysis of Section \ref{s:disturb} supports this conclusion. As Equations \ref{eq:nopps} and \ref{eq:lgicond} show, NIM will imply $-1 \leq \mean{Q_{OBS}} \leq 3 $, must hold for any ontic model, macrorealist or not.

However, in \cite{MT2012} it is argued that a weaker condition than NIM does rule out some forms of macrorealism.  This weaker condition is simply that the intervening measurement is NDM for eigenstate preparations of the macrorealist observable.  Unlike NIM, this condition may operationally verified. It is straightforward to show that quantum theory predicts $M_1$ and $M_2$ are NDM for any eigenstate preparations $\ket{1}$, $\ket{2}$ and $\ket{3}$.

The eigenstate preparation $\ket{i}$ is represented by a probability distribution over the ontic states of $\mu_i(\lambda)$.  A statistical mixture of such eigenstate preparations is represented by
\begin{equation}
\mu_M(\lambda)=p_1 \mu_1(\lambda)+p_2 \mu_2(\lambda)+p_3 \mu_3(\lambda)
\end{equation}
The macrorealist would represent a superposition of eigenstates by
\begin{equation}
\mu_S(\lambda)=p_1 \nu_1(\lambda)+p_2 \nu_2(\lambda)+p_3 \nu_3(\lambda)
\end{equation}
where $\nu_i(\lambda)>0$ only if $\xi_M(i|\lambda)=1$ for all measurements.

In \cite{MT2012} three possible approaches to macrorealism are identified:
\begin{enumerate} [MR1:]
\item $\nu_1(\lambda)$ is the same distribution over the ontic states as you would get if you prepared the system to definitely be in Box 1: $\nu_1(\lambda)=\mu_1(\lambda)$.
\item $\nu_1(\lambda)$ is a distribution over the same set of ontic states as you would get if you prepared the system to definitely be in Box 1, but is a different distribution: If $\nu_1(\lambda)>0$ then $\mu_1(\lambda)>0$ but $\nu_1(\lambda) \neq \mu_1(\lambda)$.
\item $\nu_1(\lambda)$ includes novel ontic states which do not appear when the system is prepared to definitely be in Box 1.  There exists $\lambda$ for which $\nu_1(\lambda)>0$ but $\mu_1(\lambda)=0$
\end{enumerate}
with equivalent statements for $\nu_2(\lambda)$ and Box 2 etc.

Only MR1 is incompatible with the 3 Box Paradox.  For this we will need to demonstrate that
\[
\mu(\lambda)=p_1 \mu_1(\lambda)+p_2 \mu_2(\lambda)+p_3 \mu_3(\lambda)
\]
cannot give rise to a 3 Box Paradox.

As $\mu_1(\lambda)$ is equivalent to preparing the system to be in Box 1, it corresponds to the eigenstate $\ket{1}$.  $M_1$ is operationally verified to be NDM for the eigenstate preparation $\ket{1}$.  This means that the post-measurement preparation $\mu_1^\prime(\lambda)$, given by:
\begin{equation}
 \mu_1^\prime(\lambda)= \int d\lambda_0\mu_1(\lambda_0)\xi_{M_1}(1|\lambda_0)\gamma_{M_1}(\lambda|1,\lambda_0)
\end{equation}
must be empirically indistinguishable from $\mu_1(\lambda)$.  It follows that:
\begin{eqnarray}
P_{M_1}(A,1) &=& \int d\lambda_0 p_1 \mu_1^\prime(\lambda_0)\xi_{M_A}(A|\lambda_0) \nonumber \\
 &=&\int d\lambda_0 p_1 \mu_1(\lambda_0)\xi_{M_A}(A|\lambda_0) \nonumber \\
 &=& P_{N}(A,1)
\end{eqnarray}
The Leggett Garg Inequality now follows exactly as for the case of NIM1 above.  MR1 is therefore unable to account for the statistics of the 3 Box Paradox.  Two important features may be pointed out:
\begin{itemize}
\item This rejection of MR1 does not assume any form of NIM.  The operationally verifiable properties that the measurements $M_1$ and $M_2$ are NDM for the eigenstate preparations are sufficient to rule out MR1 without appeal to NIM.
\item In 2 dimensional Hilbert spaces, a violation of the Leggett Garg Inequality may also be used to rule out MR1\cite{MT2012}.  This requires verifying two intervening measurements are NDM for eigenstate preparations, then demonstrating they are \textit{not} NDM for superposition preparations.  In the 3 Box paradox, the intervening measurements are verifiably NDM even for the superposition preparation.  Nevertheless, the Leggett Garg inequality is still violated and MR1 is ruled out.
\end{itemize}
However, we will demonstrate in Sections \ref{ss:mr3} and \ref{ss:mr2}, by explicit construction, that MR2 and MR3 are compatible with the 3 Box Paradox. 
\section{Ontic Models for the 3 Box Paradox}\label{s:ontic}
We have shown the connection of the 3 Box Paradox to measurement invasiveness and to the Leggett Garg Inequality.  We have also argued that the classical models of \cite{Kirkpatrick2003,LS2005b,RV2007} fail to reproduce a key feature of the PPS paradoxes.

We will now ask how classical can an ontic model be and yet still reproduce the 3 Box Paradox?  In particular we will be concerned with the question of macrorealism, as in the Leggett Garg analysis, and outcome definite non-contextuality, as discussed by Leifer and Spekkens\footnote{Recall from \cite{Spekkens2005a}, that in the ontic models framework, outcome definite non-contextuality means that $\xi_M(Q|\lambda)\in \{0,1\} \; \forall \lambda$ and that if $Q$ is an outcome shared by two different measurement procedures $M_1$ and $M_2$ then $\xi_{M_1}(Q|\lambda)=\xi_{M_2}(Q|\lambda) \; \forall \lambda$.  The Kochen-Specker theorem demonstrates that any ontic model for quantum theory must violate these conditions for some ontic states.}.

We exhibit three ontic models that are able to satisfy macrorealism and non-contextuality, while exactly reproducing all the relevant statistics of the 3 Box Paradox, including the NDM property of Bob's measurements.

\subsection{Alice's Cheating NIM Model}\label{ss:cheat}

The simplest way to reproduce the basic statistics of the 3 Box Paradox was implicitly suggested in Section \ref{s:disturb}, and even manages to be NIM.  Alice needs only four ontic states, $\lambda_1 - \lambda_4$, which respond deterministically and non-invasively to the measurements $M_1$,$M_2$ and $M_A$:

\[
\begin{array}{|c||cc|cc|cc|}
\hline
\multirow{2}{*}{$\xi_M(q|\lambda)$} & \multicolumn {2}{c|}{M_1} & \multicolumn {2}{c|}{M_2} & \multicolumn {2}{c|}{M_A} \\ 
           & 1 & \neg 1 & 2 & \neg 2 & A & \neg A \\ \hline \hline
\lambda_1  & 1 & 0 & 0 & 1 &  0 & 1\\
\lambda_2  & 0 & 1 & 1 & 0 & 0 & 1\\
\lambda_3 & 0 & 1 & 0 & 1 & 0 & 1\\
\lambda_4 & 1 & 0 & 1 & 0 & 1 & 0 \\
\hline
\end{array}
\]

Both measurements $M_1$ and $M_3$ are NIM.  Now the preparation:
\begin{equation}\mu_{+}(\lambda)= \frac{1}{9}\left(2 \dirdel{\lambda}{\lambda_1}+  2\dirdel{\lambda}{\lambda_2} + 4\dirdel{\lambda}{\lambda_3}+\dirdel{\lambda}{\lambda_4}  \right)
\end{equation}
successfully reproduces the statistics of the measurement sequences $(M_A)$, $(M_1,M_A)$ and $(M_2,M_A)$  exactly as in the 3 Box Paradox.  However, it fails to reproduce the statistics of $(M_1,M_2)$. The ontic state $\lambda_4$ is a cheat: Alice has simply placed a ball in Box 1 and another in Box 2.  She knows for certain that in those cases Bob will definitely see a ball if he looks in either Box 1 or Box 2, and only plays the game in those cases.

\subsection{Alice's Macrorealist Models}\label{ss:mr}
We now give the first complete description of ontic models that describe the 3 Box Paradox in which there is only one ball and which is at all times in one and only one box.  The model is outcome deterministic and measurement non-contextual.  This provides a constructive counterexample to any claim that PPS paradoxes necessarily imply contextuality\footnote{Recall that Leifer and Spekkens paper showed that if an outcome definite non-contextual model existed for a given PPS paradox, then it could not be NIM.  This did not prove that such a model actually exists!}

For this we need sixteen ontic states.  States $\lambda_1-\lambda_4$ are in Box 1, states $\lambda_5-\lambda_8$ are in Box 2 and states $\lambda_9-\lambda_{16}$ are in Box 3.

The measurement outcomes are:

\[
\begin{array} {|c||cccc|cccc|cccccccc|}
\hline
 & \multicolumn {4}{c|}{\mathrm{Box 1}} & \multicolumn {4}{c|}{\mathrm{Box 2}} & \multicolumn {8}{c|}{\mathrm{Box 3}} \\
& \lambda_1 & \lambda_2 & \lambda_3 & \lambda_4 &
\lambda_5 & \lambda_6 & \lambda_7 & \lambda_8 &
\lambda_9 & \lambda_{10} & \lambda_{11} & \lambda_{12} &
\lambda_{13} & \lambda_{14} & \lambda_{15} & \lambda_{16} \\
\hline \hline
M_1
& \multicolumn {4}{c|}{1} & \multicolumn {4}{c|}{\neg 1} & \multicolumn {8}{c|}{\neg 1} \\
M_2
& \multicolumn {4}{c|}{\neg 2} & \multicolumn {4}{c|}{2} & \multicolumn{8}{c|}{\neg 2} \\
M_A
& \neg A & A & A & \neg A & \neg A & A & A & \neg A
& \neg A & A & A & \neg A & A & \neg A & \neg A & A \\ \hline
\end{array}
\]

Bob's measurements can also change the ontic state:

\[
\begin{array} {|c||cccc|cccc|cccccccc|}
\hline
 & \multicolumn {4}{c|}{\mathrm{Box 1}} & \multicolumn {4}{c|}{\mathrm{Box 2}} & \multicolumn {8}{c|}{\mathrm{Box 3}}
\\
 & \lambda_1 & \lambda_2 & \lambda_3 & \lambda_4 &
\lambda_5 & \lambda_6 & \lambda_7 & \lambda_8 &
\lambda_9 & \lambda_{10} & \lambda_{11} & \lambda_{12} &
\lambda_{13} & \lambda_{14} & \lambda_{15} & \lambda_{16} \\
\hline \hline
M_1
& \lambda_1 & \lambda_2 & \lambda_1 & \lambda_2 &
\lambda_5 & \lambda_6 & \lambda_7 & \lambda_8 &
\lambda_9 & \lambda_{10} & \lambda_{9} & \lambda_{10} &
\lambda_{13} & \lambda_{14} & \lambda_{13} & \lambda_{14} \\
 M_2
& \lambda_1 & \lambda_2 & \lambda_3 & \lambda_4 &
\lambda_5 & \lambda_6 & \lambda_5 & \lambda_6 &
\lambda_9 & \lambda_{10} & \lambda_{11} & \lambda_{12} &
\lambda_{9} & \lambda_{10} & \lambda_{11} & \lambda_{12}
\\ \hline
\end{array}
\]

Note that none of the states change boxes as a result of the measurement.

The ontic states $\lambda_3,\lambda_4,\lambda_7,\lambda_8,\lambda_{11}-\lambda_{16}$ contain additional structure that allows them to change state when one of the boxes is tested.  As we have shown in Sections \ref{s:disturb} and \ref{s:lgi}, this disturbance is a necessary feature of any ontic model that hopes to reproduce a PPS paradox.
\subsubsection{MR3}\label{ss:mr3}
It is now straightforward to verify that the following preparations:
\begin{eqnarray}
\mu_{\ket{1}}(\lambda) &=& \frac{1}{3}\left(2 \dirdel{\lambda}{\lambda_1}+  \dirdel{\lambda}{\lambda_2}\right) \nonumber \\
\mu_{\ket{2}}(\lambda) &=& \frac{1}{3}\left(2 \dirdel{\lambda}{\lambda_5}+  \dirdel{\lambda}{\lambda_6}\right) \nonumber \\
\mu_{\ket{3}}(\lambda) &=& \frac{1}{3}\left(2 \dirdel{\lambda}{\lambda_9}+  \dirdel{\lambda}{\lambda_{10}}\right) \nonumber \\
\mu_{\ket{1+3}}(\lambda) &=& \frac{1}{6}
\left(2 \dirdel{\lambda}{\lambda_1}+ \dirdel{\lambda}{\lambda_4}
    +2 \dirdel{\lambda}{\lambda_9}+\dirdel{\lambda}{\lambda_{12}}\right) \nonumber \\
\mu_{\ket{2+3}}(\lambda) &=& \frac{1}{6}
\left(2 \dirdel{\lambda}{\lambda_5}+ \dirdel{\lambda}{\lambda_8}
    +2 \dirdel{\lambda}{\lambda_9}+\dirdel{\lambda}{\lambda_{14}}\right)\\
\mu_{\ket{1+2+3}}(\lambda) &=& \frac{1}{9}
\left(
2 \dirdel{\lambda}{\lambda_1}+  \dirdel{\lambda}{\lambda_4}
2 \dirdel{\lambda}{\lambda_5}+ \dirdel{\lambda}{\lambda_8}
    +2 \dirdel{\lambda}{\lambda_9}+\dirdel{\lambda}{\lambda_{16}}\right) \nonumber
\end{eqnarray}
produce exactly the right probabilities for all sequences of measurements: $(M_A)$, $(M_1,M_A)$, $(M_2,M_A)$, $(M_1,M_2,M_A)$, $(M_2,M_1,M_A)$, $(M_1,M_1,M_A)$, $(M_2,M_2,M_A)$ etc.

This is a macrorealist theory of type MR3: superpositions involve novel ontic states ($\lambda_4,\lambda_8,\lambda_{12},\lambda_{14},\lambda_{16}$), which do not appear in the eigenstate preparations.

\subsubsection{MR2}\label{ss:mr2}

We now exploit preparation contextuality\footnote{Preparation contextuality is the property that two distinct physical processes for preparing identical quantum states may have $\mu_1(\lambda)\neq \mu_2(\lambda)$.} to extend the previous model to show how MR2 macrorealist constructions are possible for the 3 Box Paradox: all ontic states can occur within eigenstate preparations.  To achieve this we must include the previously unused ontic states: $\lambda_3,\lambda_7,\lambda_{11},\lambda_{13}$ and $\lambda_{15}$.

Let $a_1,a_3,a,b,c>0$, and $a_1,a_3, (a+b+c)<1/3$.  These are preparation contextual parameters.  Different physical processes for preparing operational states may lead to different values of these parameters.  The following preparations include all the ontic states available, and will reproduce the correct measurement statistics for $M_1$, $M_2$ and $M_A$:

\begin{eqnarray}
\mu_{\ket{1}}(\lambda) &=& \left(\frac{2}{3}-a_1 \right) \dirdel{\lambda}{\lambda_1}+  \left(\frac{1}{3}-a_1 \right) \dirdel{\lambda}{\lambda_2}+a_1 \left(\dirdel{\lambda}{\lambda_3}+\dirdel{\lambda}{\lambda_4}\right) \nonumber \\
\mu_{\ket{2}}(\lambda) &=& \left(\frac{2}{3}-a_2 \right) \dirdel{\lambda}{\lambda_5}+  \left(\frac{1}{3}-a_2 \right) \dirdel{\lambda}{\lambda_6}+a_2 \left(\dirdel{\lambda}{\lambda_7}+\dirdel{\lambda}{\lambda_8}\right) \nonumber \\
\mu_{\ket{3}}(\lambda) &=& \left(\frac{2}{3}-(a+b+c) \right) \dirdel{\lambda}{\lambda_9}+  \left(\frac{1}{3}-(a+b+c) \right) \dirdel{\lambda}{\lambda_{10}} \\ &&
+a \left(\dirdel{\lambda}{\lambda_{11}}+\dirdel{\lambda}{\lambda_{12}}\right)
+b \left(\dirdel{\lambda}{\lambda_{13}}+\dirdel{\lambda}{\lambda_{14}}\right)
+c \left(\dirdel{\lambda}{\lambda_{15}}+\dirdel{\lambda}{\lambda_{16}}\right)  \nonumber
\end{eqnarray}

However, there is no preparation contextuality for the superposition preparations:
\begin{eqnarray}
\mu_{\ket{2+3}}(\lambda) &=& \frac{1}{6}
\left(2 \dirdel{\lambda}{\lambda_5}+ \dirdel{\lambda}{\lambda_8}
    +2 \dirdel{\lambda}{\lambda_9}+\dirdel{\lambda}{\lambda_{14}}\right) \nonumber \\
\mu_{\ket{1+3}}(\lambda) &=& \frac{1}{6}
\left(2 \dirdel{\lambda}{\lambda_1}+ \dirdel{\lambda}{\lambda_4}
    +2 \dirdel{\lambda}{\lambda_9}+\dirdel{\lambda}{\lambda_{12}}\right)\\
\mu_{\ket{1+2+3}}(\lambda) &=& \frac{1}{9}
\left(
2 \dirdel{\lambda}{\lambda_1}+  \dirdel{\lambda}{\lambda_4}
2 \dirdel{\lambda}{\lambda_5}+ \dirdel{\lambda}{\lambda_8}
    +2 \dirdel{\lambda}{\lambda_9}+\dirdel{\lambda}{\lambda_{16}}\right)  \nonumber
\end{eqnarray}
These remain the only way to reproduce the correct measurement statistics for $M_1$, $M_3$ and $M_A$ for these ontic states.  The ability to include the additional ontic states in the eigenstate preparations does not play any role in the 3 Box Paradox itself.

It should be noted here that this MR2 model is possible only with preparation contextuality for the eigenstates.  This is not necessarily the case for the 2-dimensional Hilbert space models that satisfy MR2.  So, although the 3 Box Paradox does not rule out MR2 models, it may place further constraints on what is necessary to make them succeed.

\section{Conclusion}

We have shown that PPS paradoxes necessarily violate NIM - the intervening measurements must disturb the ontic state.  We would like to emphasise that this must be the case for any ontic model for a PPS paradox, including standard quantum theory with only the wavefunction representing the ontic state. It does not just hold for ontic models involving hidden variables or which are measurement non-contextual, or outcome deterministic, or macrorealist.


We can now provide clear answers to the questions: Why would Bob have considered it reasonable to play the adversarial game?  What exactly is non-classical about the 3 Box Paradox?

If we consider the four suggestions Alice makes in Section \ref{ss:adversarial}, that convince Bob to play the game, they correspond to our analysis as follows:
\begin{enumerate}
\item The intervening measurements are NDM.  There is no obvious sense in which the measurement is leaving a mark that Alice can read.
\item Bob can verify that $P_{M_2,M_1}(2,1)=0$.  Alice's Cheating Model is ruled out.
\item If Bob is convinced that his measurement cannot introduce any disturbance to the system, he believes NIM holds, and would now be willing to play the game.  If he accepts any form of macrorealism, and simultaneously believes either NIM1 (his measurement cannot introduce any disturbance to the box he is measuring) or NIM2 (his measurement cannot introduce any disturbance to the boxes he is \textit{not} measuring) then he would also now be willing to play the game.
\item Bob can verify that his intervening measurements are NDM for eigenstate preparations.  If he believes in the MR1 form of macrorealism, he would be willing to play the game.
\end{enumerate}

The ontic models given in Section \ref{ss:mr} have not been presented with the kind of simple classical examples as Kirkpatrick's card games or Leifer and Spekken's shaken boxes.  Nevertheless they display some of the minimum requirements for a classical account of the 3 Box Paradox: the ontic state at all times is in one, and only one, of the boxes, and this does not change as a result of the measurement.  The additional classical intuitions that must be violated for the 3 Box Paradox to occur all involve measurement disturbance.  It is these properties that a classical account of the 3 Box Paradox would need to address.

The fact that the intervening measurements are NDM but cannot be NIM indicates immediately that there must be additional hidden properties to any macrorealist model.  That we can verify the intervening measurements are NDM for eigenstate preparations shows that MR1 macrorealism is ruled out, and superpositions have to be represented by novel distributions over these hidden properties.  That these hidden properties do not satisfy NIM2 is perhaps the biggest obstacle to a classical description: opening a box \textit{must} leave a record in these hidden properties even if the ball is not in that box.

The failure of MR1 and of NIM2 has already been noted in the context of the Leggett Garg Inequality in 2-dimensional Hilbert spaces.  The 3 Box Paradox displays a novel feature though: Bob's intervening measurements cannot be operationally detected.  Alice cannot infer from her measurement statistics anything about whether Bob performed an operation or not.  This is the characteristic of a true PPS paradox, and is not present in the 2-dimensional case, or in the classical examples suggested by Leifer and Spekkens or Kirkpatrick.

We now note, though, that for Bob to be sure Alice is not using the simple cheat in Section \ref{ss:cheat} we must include the following protocol in the adversarial game: Bob can choose to perform $M_1$ and then $M_2$ (or $M_2$ and then $M_1$), and wins immediately if he finds a ball in both boxes.  However, he loses if he finds only one, or no balls.  This rules out Alice's easy cheat, while Alice still wins all the time on the quantum 3 Box Paradox.

But Bob's verification that Alice is not cheating would alter the statistics of Alice's measurement.  As no-one would consider Alice's Cheating Game represents a PPS paradox, it might be argued that the combination of all three probabilistic inferences $P(1|A)=1$, $P(2|A)=1$ and $P(1 \& 2|A)=0$ is required. Verifying the third inference involves a measurement that is not NDM.  It might be argued this is no longer a true PPS paradox.  This criticism applies to all the generalisations of the 3 Box Paradox to N Boxes.  We intend to explore the further constraints raised by this additional requirement for a true PPS paradox in a future paper. 

\textbf{Acknowledgements}
I would like to thank Chris Timpson, Richard George, Erik Gauger and Andrew Briggs.  This research is supported in part by the John Templeton Foundation.

\begin{appendix}
\section{Ontic Models for PPS Games}
Re-writing the classical models for PPS paradoxes considered in \cite{Kirkpatrick2003,LS2005b,RV2007} in the ontic model formalism, shows how all involve detectable measurement disturbances.
\subsection{Kirkpatrick's Card Game}
In Kirkpatrick's card game\cite{Kirkpatrick2003}, the ontic state is represented by two piles of cards, \{Active, Passive\}, containing the following cards \{Jack of Spades, Queen of Spades, Jack of Diamonds, Queen of Diamonds, King of Hearts, King of Hearts\}.  Opening Box 1 is represented by randomly picking a card from one of the piles and asking if the Suit is Spade.  Opening Box 2 is represented by asking if the Suit is Diamond.  Alice's post-selection asks if the Face is the King.  If the measurement changes from a Suit to Face question, or vice versa, the card is picked from the Passive pile, otherwise the card is picked from Active.  After the result of the measurement, the cards are simply restored if the card was taken from Active.  If the card was taken from Passive, a new state is prepared according to the outcome of the measurement.  Five ontic states are involved:
\begin{enumerate}[${\lambda}_1$:] \setcounter{enumi}{-1}
\item Face=Q.  Active=\{QS,QD\}, Passive=\{JS,JD,2KH\}
\item Suit=S.  Active=\{JS,QS\}, Passive=\{JD,QD,2KH\}
\item Suit=D.  Active=\{JD,QD\}, Passive=\{JS,QS,2KH\}
\item Suit=not-S.  Active=\{JD,QD,2KH\}, Passive=\{JS,QS\}
\item Suit=not-D.  Active=\{JS,QS,2KH\}, Passive=\{JD,QD\}
\end{enumerate}
The outcomes and updates\footnote{The updates for measuring $M_1$ on $\lambda_4$ and $M_2$ on $\lambda_3$ follow the rules exactly as stated by Kirkpatrick.  However, it does not seem these will produce stable statistics for repeated partial measurements.} are given in the form (probability of outcome; post measurement state):
\[
\begin{array}{|c||cc|cc|cc|}
\hline
 & \multicolumn{2}{c|}{M_1} & \multicolumn{2}{c|}{M_2} & \multicolumn{2}{c|}{M_A} \\
 & 1 & \neg 1 & 2 & \neg 2 & A & \neg A \\
\hline \hline
\lambda_0 & (1/4; \lambda_1) & (3/4; \lambda_3)  & (1/4; \lambda_2) & (3/4; \lambda_4) & 0 & 1 \\
\lambda_1 & 0 & (1; \lambda_1)  & (1; \lambda_1) & 0 & 1/2 & 1/2 \\
\lambda_2 & (1; \lambda_2) & 0  & 0 & (1; \lambda_2) & 1/2 & 1/2 \\
\lambda_3 & (1; \lambda_3) & 0  & (1/2; \lambda_3) & (1/2; \lambda_3) & 0 & 1 \\
\lambda_4 & (1/2; \lambda_4) & (1/2; \lambda_4)  & 0 & (1; \lambda_4) & 0 & 1 \\
\hline
\end{array}
\]
The initial state is ``Face=Q'', from which we can quickly assess the probabilities of the sequences to be:
\[
\begin{array} {|c||c|}
\hline
M_1 & \begin{array}{cc}
        P_{M_1}(A,1) =1/8 & P_{M_1}(A,\neg 1)=0 \\
        P_{M_1}(\neg A,1)=1/8 & P_{M_1}(\neg A,\neg 1) = 3/4
     \end{array} \\
\hline
M_2 & \begin{array}{cc}
        P_{M_2}(A,2)=1/8 & P_{M_2}(A,\neg 2)=0 \\
        P_{M_2}(\neg A,2)=1/8 & P_{M_2}(\neg A,\neg 2)=3/4
     \end{array} \\
\hline
N & P_N(A)=0 \\ \hline
\end{array}
\]
While this successfully reproduces the result $P_{M_1}(1|A)=P_{M_2}(2|A)=1$, it fails to be a true PPS paradox as $P_{M_1}(A)=P_{M_2}(A)=1/8$ but $P_N(A)=0$.

Kirkpatrick's game is neither non-invasive nor macrorealist (for ontic states $\lambda_3$ and $\lambda_4$) in the sense used in this paper.  It is effectively the failure of macrorealism that Kirkpatrick argues accounts for the quantum properties.  It is possible that more complex choices of ontic states could better reproduce the 3 Box Paradox statistics.  Kirkpatrick\cite{Kirkpatrick2007} does suggest such a modification, in response to Ravon and Vaidman, so that $P_N(A) \neq 0$.

\subsection{Ravon and Vaidman's Card Game}
Ravon and Vaidman\cite{RV2007} present a simplified card game, based on Kirkpatrick's.  The number of cards are reduced, by removing the queens.
\begin{enumerate}[${\lambda}_1$:] \setcounter{enumi}{-1}
\item Face.  Active=\{\}, Passive=\{JS,JD,KH\}
\item Suit=S.  Active=\{JS\}, Passive=\{JD,KH\}
\item Suit=D.  Active=\{JD\}, Passive=\{JS,KH\}
\item Suit=not-S.  Active=\{JD,KH\}, Passive=\{JS\}
\item Suit=not-D.  Active=\{JS,KH\}, Passive=\{JD\}
\end{enumerate}
The outcomes and updates\footnote{I have had to extrapolate some of these values, as Ravon and Vaidman only specify the updates for particular measurements, instead of providing a complete set of rules. This does not impact on the measurement statistics for the actual sequences considered.  One modification allows $P_N(A)=1/3$, but still would not create a true PPS paradox.} are:
\[
\begin{array}{|c||cc|cc|cc|}
\hline
 & \multicolumn{2}{c|}{M_1} & \multicolumn{2}{c|}{M_2} & \multicolumn{2}{c|}{M_A} \\
 & 1 & \neg 1 & 2 & \neg 2 & A & \neg A \\
\hline \hline
\lambda_0 & (1/3; \lambda_1) & (2/3; \lambda_3)  & (1/3; \lambda_2) & (2/3; \lambda_4) & 0 & 1 \\
\lambda_1 & 0 & (1; \lambda_1)  & (1; \lambda_1) & 0 & 1/2 & 1/2 \\
\lambda_2 & (1; \lambda_2) & 0  & 0 & (1; \lambda_2) & 1/2 & 1/2 \\
\lambda_3 & (1; \lambda_3) & 0  & (1/2; \lambda_3) & (1/2; \lambda_3) & 0 & 1 \\
\lambda_4 & (1/2; \lambda_4) & (1/2; \lambda_4)  & 0 & (1; \lambda_4) & 0 & 1 \\
\hline
\end{array}
\]
the initial state is ``Face'', giving the probabilities:
\[
\begin{array}{|c||c|} \hline
M_1 & \begin{array}{cc}
        P_{M_1}(A,1)=1/6 & P_{M_1}(A,\neg 1)=0 \\
        P_{M_1}(\neg A,1)=1/6 & P_{M_1}(\neg A,\neg 1)=2/3
    \end{array} \\ \hline
M_2 & \begin{array}{cc}
        P_{M_2}(A,2)=1/6 & P_{M_2}(A,\neg 2)=0 \\
        P_{M_2}(\neg A,2)=1/6 & P_{M_2}(\neg A,\neg 2)=2/3
    \end{array} \\ \hline
N & P_N(A)=0 \\ \hline
\end{array}
\]
Again, the PPS paradox occurs $P_{M_1}(1|A)=P_{M_2}(2|A)=1$, but it fails to be a true PPS paradox as $P_{M_1}(A)=P_{M_2}(A)=1/6$ but $P_N(A)=0$.

\subsection{Leifer and Spekken's Ball Game}
Leifer and Spekkens\cite{LS2005b} consider a ball within a square box.  The ball may be in one of four positions: top  left; top right; bottom left; and bottom right.  The box may be divided into two compartments: either top-bottom or left-right.  The location of the ball may only be measured by dividing the box into two compartments and shaking one of the compartments.  A rattling sound indicates the ball is present but disturbs it.  No rattle indicates the ball is in the other compartment but does not disturb it.  They consider preparing the ball to be in the bottom.  Then a measurement is made of either the left or right compartment, finally followed by a post-selection on a successful top measurement.

\begin{enumerate}[${\lambda}_1$:]
\item Bottom left
\item Bottom right
\item Top left
\item Top right
\end{enumerate}
The outcomes and updates are:

\[
\begin{array}{|c||cc|cc|cc|}
\hline
 & \multicolumn{2}{c|}{M_L} & \multicolumn{2}{c|}{M_R} & \multicolumn{2}{c|}{M_T} \\
 & L & \neg L & R & \neg R & T & \neg T \\
\hline \hline
\lambda_1 & (1; 1/2(\lambda_1+\lambda_3)) & 0  & 0 & (1; \lambda_1) & 0 & (1; \lambda_1)\\
\lambda_2 & 0 & (1; \lambda_2) & (1; 1/2(\lambda_2+\lambda_4)) & 0 &  0 & (1; \lambda_2) \\
\lambda_3 & (1; 1/2(\lambda_1+\lambda_3)) & 0  & 0 & (1; \lambda_3) & 0 & (1; 1/2(\lambda_3+\lambda_4))\\
\lambda_4 & 0 & (1; \lambda_4) & (1; 1/2(\lambda_2+\lambda_4)) & 0 &  0 & (1; 1/2(\lambda_3+\lambda_4)) \\
\hline
\end{array}
\]

The system is initially prepared in the state $1/2(\lambda_1+\lambda_2)$.  The probabilities of the sequences are now:
\[
\begin{array}{|c||c|} \hline
M_L & \begin{array}{cc}
        P_{M_L}(T,L)=1/4 & P_{M_L}(T,\neg L)=0 \\
        P_{M_L}(\neg T,L)=1/4 & P_{M_L}(\neg T,\neg L)=1/2
    \end{array} \\ \hline
M_R & \begin{array}{cc}
        P_{M_R}(T,R)=1/4 & P_{M_R}(T,\neg R)=0 \\
        P_{M_R}(\neg T,R)=1/4 & P_{M_R}(\neg T,\neg R)=1/2
    \end{array} \\ \hline
N & P_N(T)=0 \\ \hline
\end{array}
\]
The PPS paradox occurs as $P_{M_L}(L|T)=P_{M_R}(R|T)=1$, but it fails to be a true PPS paradox as $P_{M_L}(T)=P_{M_R}(T)=1/4$ but $P_N(T)=0$.  Unlike Kirkpatrick's model, both macrorealism, in the strong MR1 form, and $NIM2$ are guaranteed to hold.   It follows that no possible modification of this classical model could simulate the 3 Box Paradox or violate the Leggett Garg Inequality. \end{appendix}
\bibliographystyle{alpha}

\begin{thebibliography}{GRM{\etalchar{+}}12}

\bibitem[AV91]{AV1991}
Y~Aharonov, , and L~Vaidman.
\newblock Complete description of a quantum system at a given time.
\newblock {\em Journal of Physics A}, 24:2315--2328, 1991.

\bibitem[BGG94]{BGG1994}
F~Benatti, G~Ghirardi, and R~Grassi.
\newblock On some recent proposals for testing macrorealism versus quantum
  mechanics.
\newblock {\em Foundations of Physics Letters}, 7(2):105--126, 1994.

\bibitem[Cli91]{Clifton1990}
R~K Clifton.
\newblock Noninvasive measurability, negative result measurements, and
  watched-pots.
\newblock In P~Lahti and P~Mittelstaedt, editors, {\em Symposium on the
  Foundations of Modern Physics 1990}. World Scientific, Singapore, 1991.

\bibitem[Coh95]{Cohen1995}
O~Cohen.
\newblock Pre- and postselected quantum systems, counterfactual measurements,
  and consistent histories.
\newblock {\em Physical Review A}, 51:4373--4380, 1995.

\bibitem[GRM{\etalchar{+}}12]{GMR+2012}
R~E George, L~Robledo, O~J~E Maroney, M~Blok, H~Bernien, M~L Markham, D~J
  Twitchen, J~J~L Morton, G~A~D Briggs, and R~Hanso.
\newblock Opening up the quantum three-box problem with undetectable
  measurements.
\newblock {\em ArXiv e-print service}, 2012.
\newblock arXiv.org://1205.2594.

\bibitem[HR07]{HR2007}
N~Harrigan and T~Rudolph.
\newblock Ontological models and the interpretation of contextuality.
\newblock {\em ArXiv e-print service}, 2007.
\newblock arxiv.org://quant-ph/0709.4266.

\bibitem[HS07]{HS2007}
N~Harrigan and R~W Spekkens.
\newblock Einstein, incompleteness and the epistemic view of quantum states.
\newblock {\em ArXiv e-print service}, 2007.
\newblock arxiv.org://quant-ph/0706.266.

\bibitem[Kas03]{Kastner2003}
R~E Kastner.
\newblock The nature of the controversy over time-symmetric quantum
  counterfactuals.
\newblock {\em Philosophy of Science}, 70:145--163, 2003.

\bibitem[Kir03]{Kirkpatrick2003}
K~A Kirkpatrick.
\newblock Classical three box paradox.
\newblock {\em Journal of Physics A}, 36:4891--4900, 2003.

\bibitem[Kir07]{Kirkpatrick2007}
K~A Kirkpatrick.
\newblock Reply to `{T}he three-box paradox revisited' by {T R}avon and {L
  V}aidman.
\newblock {\em Journal of Physics A}, 40:2883--2890, 2007.

\bibitem[KSY{\etalchar{+}}11]{KSY+2011}
P~Kolenderski, U~Sinha, L~Youning, T~Zhao, M~Volpini, A~Cabello, R~Laflamme,
  and T~Jennewein.
\newblock Playing the {A}haronov-{V}aidman quantum game with a {Y}oung type
  photonic qutrit.
\newblock {\em ArXiv e-print service}, 2011.
\newblock arXiv.org://quant-ph/1107.5828.

\bibitem[LG85]{LG1985}
A~J Leggett and A~Garg.
\newblock Quantum mechanics versus macroscopic realism: Is the flux there when
  nobody looks?
\newblock {\em Physical Review Letters}, 54(9):857--860, 1985.

\bibitem[LS05a]{LS2005b}
M~S Leifer and R~W Spekkens.
\newblock Logical pre- and post-selection paradoxes, measurement-disturbance
  and contextuality.
\newblock {\em International Journal of Theoretical Physics}, 44:1977--1987,
  2005.
\newblock arxiv.org://quant-ph/0412179.

\bibitem[LS05b]{LS2005a}
M~S Leifer and R~W Spekkens.
\newblock Pre- and post-selection paradoxes and contextuality in quantum
  mechanics.
\newblock {\em Physical Review Letters}, 95:200405, 2005.

\bibitem[MT12]{MT2012}
O~J~E Maroney and C~G Timpson.
\newblock What exactly does a violation of the {L}egget-{G}arg {I}nequality
  actually imply?
\newblock {\em forthcoming}, 2012.

\bibitem[RLS04]{RLS2004}
K~J Resch, J~S Lundeen, and A~M Steinberg.
\newblock Experimental realization of the quantum box problem.
\newblock {\em Physics Letters A}, 324:125, 2004.

\bibitem[Rud06]{Rudolph2006}
T~Rudolph.
\newblock Ontological models for quantum mechanics and the {K}ochen-{S}pecker
  theorem.
\newblock {\em ArXiv e-print service}, 2006.
\newblock arxiv.org://quant-ph/0608.120.

\bibitem[RV07]{RV2007}
T~Ravon and L~Vaidman.
\newblock The three-box paradox revisited.
\newblock {\em Journal of Physics A}, 40:2873--2882, 2007.

\bibitem[SGM08]{SPS2008}
D~Sokolovski, I~Puerto Gimenez, and R~Sala Mayato.
\newblock Path integrals, the {ABL} rule and the three-box paradox.
\newblock {\em Physics Letters A}, 372:6578--6583, 2008.

\bibitem[Spe05]{Spekkens2005a}
R~W Spekkens.
\newblock Contextuality for preparations, transformations and unsharp
  measurements.
\newblock {\em Physical Review A}, 71:052108, 2005.

\bibitem[SS93]{SS1993}
W~D Sharp and N~Shanks.
\newblock {\em Philosophy of Science}, 60:488, 1993.

\end{thebibliography}
\newcommand{\etalchar}[1]{$^{#1}$}

\end{document}